# Composable Industrial Internet Applications for Tiered Architectures

K. Eric Harper, Karen Smiley
ABB Corporate Research
940 Main Campus Drive
Raleigh, NC USA
{eric.e.harper, karen.smiley} @ us.abb.com

Thijmen de Gooijer
ABB Corporate Research
Forskargränd 7
721 78 Västerås, Sweden
thijmen.de-gooijer @ se.abb.com

*Abstract* – A single vendor cannot provide complete Industrial Internet of Things (IIoT) end-to-end solutions because cooperation is required from multiple parties. Therefore, interoperability is a key architectural quality. Composability of capabilities, information and configuration is the prerequisite for interoperability, supported by a data storage infrastructure and defined set of service interfaces to build applications. Secure collection, transport and storage of data and algorithms are expectations for collaborative participation in any IIoT solution. Participants require control of their data ownership and confidentiality. We propose an Internet of Things, Services and People (IoTSP) application development and management framework which includes components for data storage, algorithm design and packaging, and computation execution. Applications use clusters of platform services, organized in tiers, and local access to data to reduce complexity and enhance reliable data exchange. Since communication is less reliable across tiers, data is synchronized between storage replicas when communication is available. The platform services provide a common ecosystem to exchange data uniting data storage, applications, and components that process the data. Configuration and orchestration of the tiers are managed using shared tools and facilities. The platform promotes the data storage components to be peers of the applications where each data owner is in control of when and how much information is shared with a service provider. The service components and applications are securely integrated using local event and data exchange communication channels. This tiered architecture for composable applications reduces the cyber security attack surface and enables individual tiers to operate autonomously, while addressing key interoperability concerns. We present our framework using predictive maintenance for power transformers as an example, and evaluate compatibility of our vision with an emerging set of standards.

## I. INTRODUCTION

Our society has broadly adopted mobile smart phone technology over the past years. With it comes a familiar digital ecosystem including wireless connectivity, app store content deployment, and different platform and application development frameworks. Platform providers are battling over control of the ecosystem because of the advantages and profits that accrue from market adoption. We see these developments repeating in the Industrial Internet of Things (IIoT) domain.

Key IIoT vendors are partnering to deliver end-to-end systems that collect industrial data, perform calculations, and use the results to enhance current operations. Their offerings create entirely new businesses, such as advanced fleet analytics and predictive maintenance. However, each vendor's success requires cooperation from multiple parties. Factory automation vendors do not have the skills and resources to scalably and securely manage large data centers. Cloud hosting and mobile computing vendors lack the industrial expertise to design and configure applications for IIoT scenarios. Each of the vendors and customers in an end-to-end solution form an ecosystem, as illustrated in Figure 1. Aggregating and sharing the information using interoperability brings the promised IIoT benefits and actionable insights.

Multiple benefits come from providing end-to-end solutions for IIoT. First, standardizing access to platform capabilities and information accelerates the integration process for existing customer deployments and third parties, rather than starting from scratch with each vendor. Next, simplifying the interfaces to just what is needed reduces the level of expertise needed to complete the integration tasks. Finally, common integration techniques are easier to manage and monitor for proper operation.

Based on these insights, we propose an Internet of Things, Services and People (IoTSP) application development and management framework that addresses data ownership, composability and interoperability concerns. IoTSP extends IIoT and is the next logical step in the evolution of industrial automation. *Things* are industrial assets equipped with sensors, actuators, computing power and software. New *Service* models leverage this technology and ecosystem to turn identified improvements into actions. *People* program and control all processes and activities performed by things, and benefit from the resulting value. The framework includes components for data storage, algorithm design and packaging, and computation execution. The core ideas of our framework and how it is configured are described in Sections III, IV and V.

Before defining our framework in detail, we describe industrial concepts relevant to our vision and reference related work in Section II. Later we discuss how select IIoT standards support our ideas in Section VI. We round off the paper with our conclusions and contemplations for future work in Section VII.

## II. BACKGROUND AND RELATED WORK

### A. Background

For most IIoT vendors, delivering digital business value starts with a single compelling software application for a small set of customers. With that success, more customers are added who can share the same benefits. It is rare that an application operates in a vacuum. Other segments of a customer's business may also be supported by digital technology, leading to expectations that internal data is shared and workflows automate the business processes. A better-organized vendor active in the customer environment can dictate the terms of those integrations in the absence of standards or common best practices.

Figure 1 shows an example end result from this evolution. Multiple applications can share components and data, and create savings that are passed on to customers in competitive markets. Common frameworks and ecosystems increase speed to market, reduce maintenance costs, and lower development risks when extensions are needed.

One standard for organizing enterprise and control systems is ANSI/ISA-95 [1]. This hierarchical architecture defines the integration from the Physical Plant to External Systems with five Levels. Information technology (IT) and operational technology (OT) meet and are integrated in Level Three. Originally designed for on-premise systems, the standard does not exclude deployments of Level Two and above to be in hosted or cloud platforms.

A robust ecosystem for IIoT will facilitate collaboration across different businesses and vendors. People provide the knowledge and subject matter expertise to make business decisions. Detailed awareness of the current process conditions is the first step to empower those decisions. The communication goes both ways: machine to human and vice versa. In all cases the content must be secure so that only authorized participants have access.

Representation of machine data depends on a variety of factors, including existing technology stacks, previous integrations and performance concerns. Support for existing formats must be preserved in the ecosystem and potentially translated for conformance with other environments. Application processing benefits from common data formats that enable consistent computations and interpretation of results.

Application integration provides access to the right information in a secure and timely manner. Some operations can be automated, while others require people to validate the recommended actions. First, the ecosystem orchestrates data analysis and business decision collaboration to improve safety, prevent failures and reduce downtime. Next, tools support an environment where third party Subject Matter Experts (SMEs) can contribute their process and equipment knowledge and experience. Third, the infrastructure protects intellectual property associated with data and algorithms to address customer objections for participation. Finally, the configuration enables data owners to define and provision a finite set of policies for data access and retention that span the expected integration scenarios and standardize expectations for privacy and security.

Our previous work on industrial analytics pipelines [2] identifies that designing a custom architecture for each application is expensive, and proposes a framework to develop industrial applications. Our pending work on data management, ownership and access control [3] introduces the idea of microdatabases, allowing a variety of information models similar to how microservices encourage fit to purpose APIs (Application Programming Interface).

### B. Related Work

#### 1) IIoT Standards Consortia

Consortia are organizing industry stakeholders to help define and shape the next generation of industrial automation technology. Industrie 4.0 provides a Reference Architecture Model [6] that recommends specific standards and focuses on OT manufacturing productivity, with the expectation that the ideas will have broader applicability for additional application scenarios. On the other hand, the Industrial Internet Consortium Industrial Internet Reference Architecture [7] covers the entire range of IIoT use cases from primarily an IT perspective.

The European Research Cluster on the Internet of Things, IERC [12], coordinates and builds a broad-based consensus for realization of the Internet of Things vision. The effort has a number of deliverables, including a catalog of IoT naming, addressing and discovery schemes [13]. The schemes include solutions based on legacy standards, semantic solutions that rely on emerging standards, as well as a federated approach for interoperability of legacy and emerging solutions. Another deliverable [14] catalogues future technology developments and research needs.

These IIoT consortia focus on concepts, architecture and strategy. Several academics note this trend and delve deeper

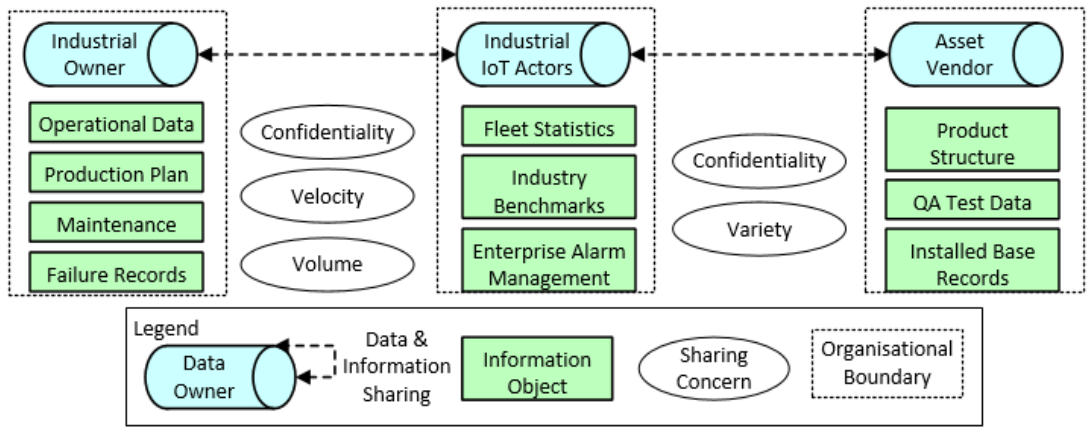

Figure 1. Industrial IoT Collaboration

[15] to understand what is novel about IIoT. Digitalization and the Internet are not new, and interconnected physical systems, mobile information and expectations for autonomous decisions have the potential to speed up the pace of industry. A group of standards researchers enhance the consortia efforts by organizing the industrial ecosystem as three dimensions: product, production and business [16]. Their opinion is that "… tighter integration within and across the three dimensions will result in faster product-innovation cycles, more efficient supply chains, and more flexibility in production systems."

*2) Distributed Application Frameworks*

The Topology and Orchestration Specification for Cloud Applications (TOSCA) [4] defines a portability framework for migration of applications across cloud services. Our work differs from TOSCA as we do not limit our scope to cloud technologies for software operation, and we include industrial control platforms as well as traditional enterprise IT.

The Standards for M2M and the Internet of Things (oneM2M) [5] functional architecture provides common point-to-point communication between tiers and enhances cross-vendor integration possibilities. An implementation of oneM2M could provide the required technology for the cross-tier communication our work assumes, including application deployment capabilities. However, oneM2M does not prescribe how to achieve cross-tier application-level collaboration and deployment or data sharing paradigms.

Open Operations & Maintenance (OpenO&M) [8] is a joint initiative of The International Society of Automation (ISA) [9], Machinery Information Management Open System Alliance (MIMOSA) [10], Manufacturing Enterprise Solutions Association International (MESA), OPC Foundation, and Open Applications Group (OAGi). The Open O&M framework encompasses interoperability standards for exchanging O&M data and associated context. MIMOSA's Common Interoperability Register (CIR) supports data fusion across systems which use different identifiers for the exact same object. The MIMOSA Open Object Registry supports a "full mesh network" for maintaining interrelationships in a Services Oriented Architecture among people, processes, and systems.

The Open Services Gateway initiative (OSGi) [11] focuses on the Java technology stack and devices at the network edge to deliver an open, common, modular architecture. Their Service Layer simplifies the development and deployment of service bundles by decoupling service specifications from implementations.

A survey of the past, current and future integration of distributed enterprise applications [17] reveals increasing use of web standards. The authors raise concerns regarding security risks because data exchange transits multiple networks. Additional researchers propose that Service-Oriented Architectures (SOA) can be used for automation industry applications [18]. The authors' key principles are reusability, contracts, loose coupling, abstraction, composability, autonomy, statelessness, and discoverability.

Researchers immersed in the IoT (with potentially billions of devices) recognize a crucial need to minimize, if not eliminate, the need for human intervention for the configuration of newly deployed objects [19]. They propose a scalable and self-configuring architecture for service and resource discovery using the Constrained Application Protocol (CoAP). Other contributors envision tools that allow SMEs to retrieve the data they want using semantic-driven models, without knowing the underlying technical details of the sensors and data processing components [20]. The tools enable configuration for orchestration of data ingest and processing, context discovery, and quality trade-offs.

A comparative study of web development technologies finds all of the approaches leverage the Model-View-Controller (MVC) pattern [21]. Our work extends this applicability to the Industrial Internet where the model represents the asset types and collected data, the view is provided by the app service implementations, and the controller is realized by the orchestrating app. Other researchers concur, proposing a novel approach for programming applications across 3-tiers using a distributed extension of the MVC architecture [22]. Different in our vision, the MVC pattern interactions do not cross tier boundaries.

*C. Transformer Case Study*

Throughout this paper, the proposed architecture is illustrated by reference to an example application which assesses the health of industrial assets, such as a power transformer. Transformers in utility substations are a significant capital investment and need to be continuously operated over a number of decades with minimal outages. A transformer asset health algorithm encapsulates the deep knowledge and analysis experience of a technical expert with operational performance of power transformers, how and why they fail or degrade, and the cost-effectiveness of various repair or refurbishment alternatives. The algorithm applies to power transformers produced in a broad range of voltage classes by a variety of vendors, and with different types of insulation, protection, and built-in or accessory sensors and monitors. A transformer fleet asset health application, in a different tier, aggregates and provides additional analysis using the results of the algorithm.

III. ECOSYSTEM

The IoTSP ecosystem and motivations for a composable development and management framework can be understood by examining the participants in the ecosystem. Figure 2 shows the stakeholders that have interest. First, from a traditional automation perspective, the Industrial Owner has responsibility for the equipment. In our example, the electric power utility operates substation transformers as assets to convert from transmission to distribution voltages. The transformers are provided by the Asset Vendor, with deep knowledge of how the equipment is designed and manufactured. The Service Vendor understands the necessary conditions for safe and efficient operation of the assets.

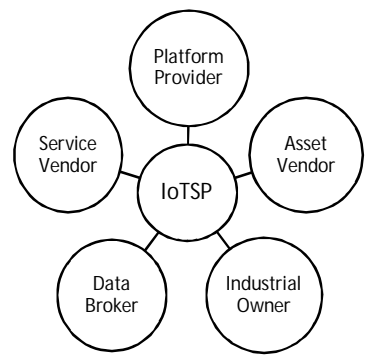

Figure 2. IoTSP Stakeholders

The digital aspects of IoTSP are orchestrated by the Platform Provider. Going back to our example, transformer conditions, both electrical and mechanical, are collected as historical records within the Industrial Owner's organization, and then shared with the Asset and Service Vendors in hosted solutions implemented using the Platform Provider's technology. The collected readings are fused with other information provided by a Data Broker, for example the weather conditions at the substation. Business value is created by enabling supervision and analysis of the transformer conditions by all stakeholders.

IoTSP configuration depends on a number of factors, including the types of assets, network connections, service contracts, regulatory oversight, and expectations regarding data ownership. All the stakeholders contribute to the configuration. Similarly, computations are orchestrated and performed to manage the assets and the IT systems. Asset Vendors provide algorithms that map equipment conditions to first principles models, for example the thermodynamics of transformer cooling. Service Vendors review historical records and anticipate the need for scheduled or pre-emptive maintenance. Industrial Owners optimize asset operation by computing and monitoring key performance indicators (KPIs), such as how often the transformer has been overloaded. Platform Providers review data processing capacity, latencies and throughput and adjust the resources as needed.

IoTSP is represented by a number of tiers, each able to operate autonomously or in a degraded fashion based on the available data and service implementations. Not all tiers are necessary. One example set of tiers is shown in Figure 3. Tiers close to devices respond immediately to meet mission critical requirements, for example a temperature sensor on a transformer that triggers an alarm; whereas the higher level tiers are better at collecting data over time and using the historical records for planning and driving longer term business decisions. There can be multiple global cloud tiers, one for each vendor in the IoTSP ecosystem (perhaps there are several different brands of transformers in a grid). Regional tiers may be required due to country-specific regulations for data sharing outside of jurisdictions. Local and plant tiers occur naturally as legacy operational technology deployments, and device tiers arise as embedded computers expand their storage and computing capacities. The Plant tier represents any industrial installation, not just manufacturing.

Automation systems integrate data at the local level, providing access to readings and computing key performance indicators across the processes in a plant. This enables accurate and timely decision support without exposure of process intellectual property. On the other hand, sharing filtered plant data in regional or global tiers, for example all the transformers in a state or country, enables data fusion and fleet analysis beyond the local plant context. The resulting experiential model is brought back to all the plants to detect and manage abnormal conditions.

In a similar way, enterprise business systems integrate data locally and are able to operate autonomously to increase productivity and reduce costs. Raw and aggregated values are collected, and key performance indicators calculated to facilitate and regulate operations, as well as provide decision support. Businesses hierarchically integrate multiple systems in regional tiers to guide workforce scheduling and organize logistics.

Data ownership and interoperability expectations have a strong influence over the architecture and configuration of the Industrial Internet. As shown in Figure 4, IoTSP is characterized by both high interoperability and a high expectation of data ownership preservation, regardless of where the information is copied. In contrast, mobile applications regularly require personal consent to use the data on your phone but rarely connect with other applications. Enterprise applications integrate with third party data sources and tools, but the ecosystem does not extend beyond the boundaries of the organization. Finally, IoT applications typically create closed environments where data ownership is lost to the service provider as the readings are transferred to the cloud.

## IV. TIERED ARCHITECTURE

### A. Tiers

We define a tier as a collection of software elements that have access to co-located resources and data. With co-location there is reliable communication between the elements, but intermittent communication between tiers. Data transfer between tiers does not happen in real time. This layered architecture pattern ensures low latency access to services and data, accepting the trade-off of computing with potentially stale state that has not been updated by another tier. This is a reasonable compromise if processing is performed on a cyclic basis, guaranteeing eventual consistency of data and results.

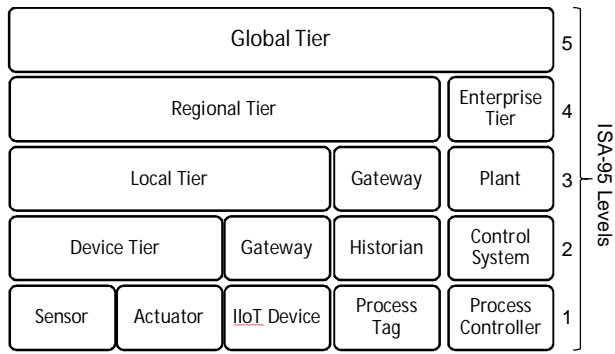

Figure 3. IoTSP Example Tiers

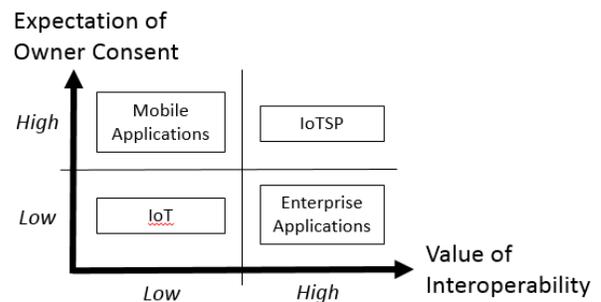

Figure 4 Industrial Internet Critical Factors

A tier serves as a security domain, indicating where the trusted network ends. Figure 1 and Figure 3 illustrate ideas for possible tier splits from different perspectives: organizational borders, and locality. An ecosystem consists of a flexible number of tiers: the number of tiers is not predefined. A tier boundary could coincide with a network segment or autonomous system in an IP network. A single operator (e.g. company) controls every software artifact within its tier.

*B. Microdatabases*

Our microdatabase concept applies microservice paradigms to data storage and access. Microservices are small, autonomous services focused on doing one thing well [23]: accepting the component complexity risk to enable better development productivity and faster time to market. Microdatabases are configured with targeted information models allowing diversity in data representation and relationships. This parallels the trend in microservices where every service implementation has a unique set of APIs specific to the provided capabilities, fit for purpose, and applications must know how to use them. In a similar way, the information model enables discovery and classification (tagging) of asset types, properties and instances.

A microdatabase is a container for a set of column stores. The API client interfaces for microdatabases are shown in Figure 5. Each client must authenticate before gaining access. The Information Model client navigates configured asset types and properties of their instances, returning discovered identifiers used by Data Exchange clients to access key-value pairs in column stores via Create, Read, Update, and Delete (CRUD).

Each microdatabase serves as a publish and subscribe hub in its tier, generating notifications associated with data exchange operations. For example, adding a new transformer temperature reading publishes an event. A Subscription client registered for the associated column store receives the notification and can perform the necessary Read to query for the new value. Only microdatabases publish events to prevent event cycles, and neither applications nor service implementations can use notifications for data exchange.

Similarly configured microdatabases can be deployed in adjacent tiers. The Replication client bi-directionally transfers records between column stores according to filter criteria defined by the data owner. This communication between tiers ensures data stays within a trusted ecosystem and reduces the cyber security attack surface. Customers may be wary of applications that advertise use of other data sharing mechanisms.

The microdatabase owner controls the access privileges using the Admin client. Microdatabases are deployed from templates and then configured by the data owner. A microdatabase imposes a security domain to protect and manage access to data. Data owners choose (from a pre-defined set) the End User License Agreement (EULA) policies by which sharing is allowed, protecting intellectual property and sensitive information. Synchronized replicas in adjacent tiers are guarded by the same controls. The EULA applies to the entire microdatabase and policies are machine readable to assist with automatic configuration for each column store.

One special use of a column store is to manage work requests across tiers. Work requests written to a microdatabase in the local tier replicate with other tiers, triggering remote activities whose results are collected and replicated back to the requesting tier. These mechanisms can adapt to intermittent connectivity between tiers.

Within a tier, the microdatabase data owner authorizes applications and service implementations to read information and to receive events. The EULA and sharing policy impact the service and application functionality available. For example, to receive fleet or industry benchmarking from a vendor or broker, the customer must share her own data; alternatively, the customer can share less or only summarized data and accept reduced access to service benchmarks.

*C. Platform and Application Services*

Microdatabases enable the interactions in a tier, providing platform services supporting role-based access control for data exchange and notification.

Figure 6 shows a logical view of the platform. App services query column stores in the microdatabases, caching, fusing and processing the data. The app services are integrated by apps from the App Store, where each component has a specific role and responsibilities. These components serve as composable building blocks for applications, with processing results stored into columns stores in the microdatabases.

Not all component interactions are with microdatabases. App services can integrate with any resources on a tier, enabling interactions with related data interfaces of legacy systems. The interfaces for standard functionality offered by the IoTSP platform services are consistent across tiers and versioned. Each tier must agree on one consistent version, however different tiers may run different versions, and not all platform service operations will be available at all tiers. Additional platform service components include an interoperability register, user identity management, and certificate store.

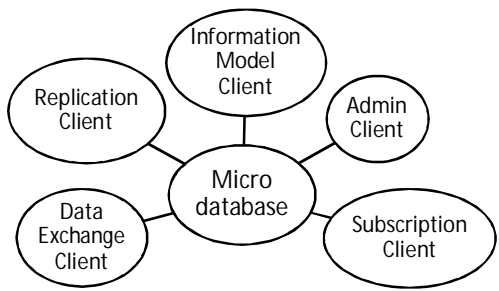

Figure 5 IoTSP Microdatabase Context

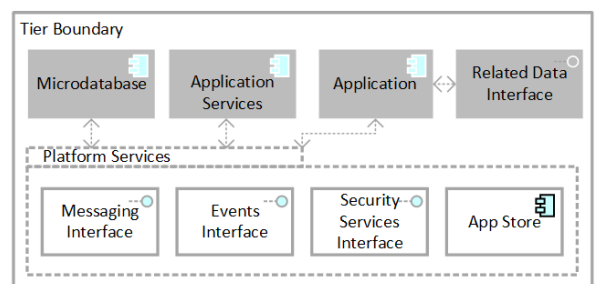

Figure 6. Framework architecture components within a tier

### D. Applications

Data assets are at the core of our framework, but it is the applications that deliver actual value to vendors and industrial asset owners. Applications compose microdatabases with the orchestration of service implementations. Examples are data analytics for failure detection or condition based maintenance, collaborative troubleshooting and asset configuration optimization.

The functionality of an application for an industrial asset owner can span multiple different tiers, as shown in Figure 7. SLAs, data availability and access controls drive the orchestration of analytics applications. Best practice for an app and its services is to interact only with resources in its tier, because the customer could be disconnected. The framework however does not prevent applications from connecting to available URLs to access data not stored in a microdatabase.

The preferred mechanism for applications to interact across tiers is through the microdatabases and their replication mechanism. Interacting apps and services require configuration of microdatabases in their respective tiers to act as message routers and brokers to interconnect application components. Using the microdatabases as intermediaries ensures enforcement of cross tier data sharing policies. As described for microdatabases, applications can request work from other applications in the same or other tiers by adding a work request in a designated column store. The microdatabase then checks the permissions and publishes an event or hands-over the work request to a microdatabase on another tier.

### E. Model-View-Controller

The MVC pattern is most familiar in client application designs, especially the Spring framework for Java [24]. The same pattern can be used for web service-based applications, as shown in Figure 8. The arrows show information flow between the elements. The microdatabases realize the Model elements of the pattern using APIs for data exchange and notification. The services implement View elements of the pattern, accessing, processing and fusing the data. The applications (Apps) realize the Controller elements of the pattern to orchestrate the services based on scheduled and ad-hoc work requests. The MVC pattern is implemented within a tier.

The transformer example can be realized using a MVC pattern. First, the sensor data is collected and ingested into column stores in the microdatabase. The microdatabase is configured with the information model for the transformer, making it possible to discover the different properties and access their values in the associated column stores. The algorithms to interpret the data are encapsulated within services deployed in the same tier as the microdatabase. When new values are added to the microdatabase, the transformer analysis app and app services are notified of the updates, the app invokes the transformer algorithm app service. The service writes the analysis results to column store in the microdatabase, which notifies the app the analysis is complete.

### F. App Stores

The app store provides a catalogue of all vendor and 3$^{rd}$ party applications and microdatabases. Distribution through the app store enables consistency, safeguards authenticity, and enforces security reviews. Furthermore, version management and roll-out of updates are simplified and traceable using the app store pattern prevalent in the mobile smart phone world.

We differ from mobile app stores in that we distribute not only applications but also microdatabases through the app store. We consider these merely special applications that have data management as their chief role. Vendors, including third parties, can deploy microdatabases using the app store, with provisions for customers to wire them to data sources using configuration tools.

### G. Related Data Interface

Not all the data an application requires may be offered through a microdatabase interface. A clear case for this need is the presence of legacy systems that cannot be wrapped to provide the microdatabase interfaces. We refer to any other non-framework interface as a related data interface. Database systems realizing related data interfaces may or may not respect tier boundaries (see Figure 1). This implies that applications may be able to connect across tiers using such interfaces. The app store provides transparency to the tier owner as to which connections an application depends upon. It is up to the tier owner to decide which applications to deploy and which privileges to grant the application, and to protect the tier boundaries to ensure the desired data confidentiality.

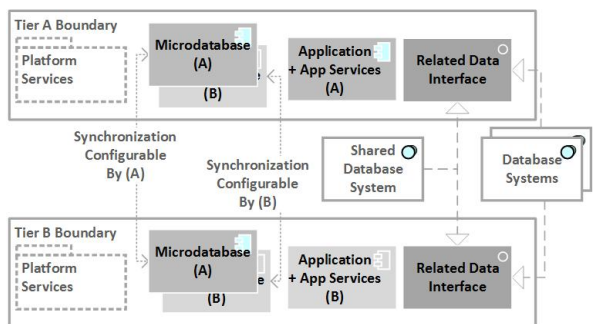

Figure 7. Cross Tier Interactions

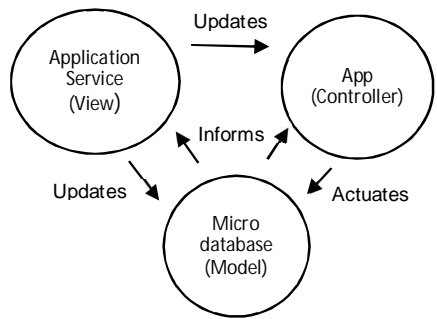

Figure 8. IoTSP Model-View-Controller

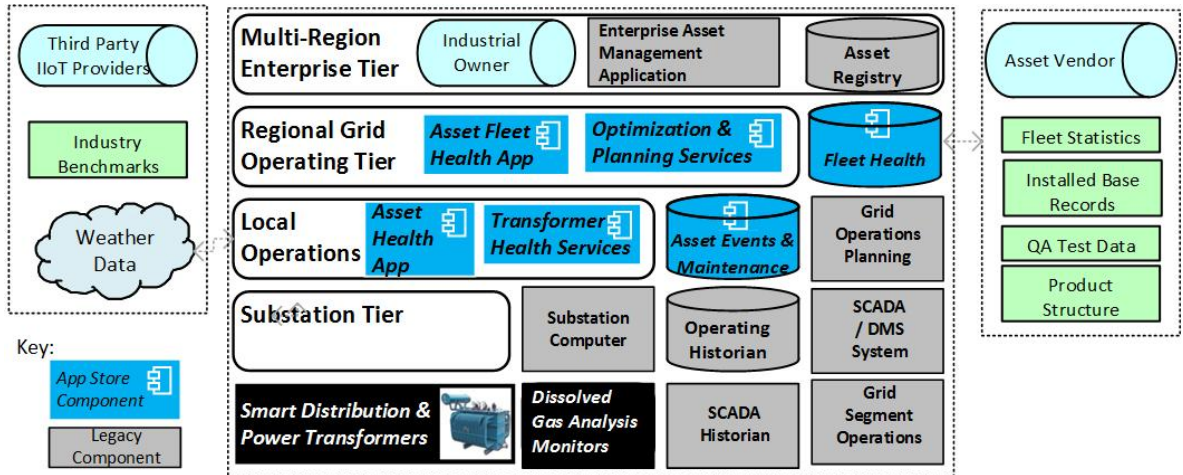

Figure 9. IoTSP Example Case Study mapped to Tiers and Collaborators

## V. APPLICATION CONFIGURATION AND ENGINEERING

IIoT requires sensor-level and system-level configurations, and for our example we focus on system-level configuration. SMEs participate in three roles to contribute ecosystem elements to the app store and in configure them [27]: *Technical SMEs*, expert at operational performance of industrial domains and operating assets, how and why they fail or degrade, and the cost-effectiveness of diagnosis, repair, and refurbishment; *Solution SMEs*, familiar with end user and customer needs and digital technologies; and *Integration SMEs*, who understand the complexities of customer operating environments and toolset configurations. A single person might fill more than one role.

Building a composable application instance begins with a vendor Solution SME selecting components from the app store. Each component is deployed to a tier, configured, and then validated and possibly tuned, by a Solution SME and/or Integration SME. Three sets of configurations are needed: for microdatabases to load data (e.g., tags or signals) from the sources which generate them, to accept calculated values for storage, and to share their data; for app services and apps to associate their inputs and outputs with microdatabases or related data interfaces; and for app services and apps to fetch and store data to microdatabases and issue work requests.

Specific scenarios for configuring and managing these composable applications will be discussed with respect to Figure 9, which shows our example case study overlaid on the tier example from Figure 3 and the IoTSP collaborators from Figure 1. Technical SMEs with deep knowledge of power and distribution transformers contribute transformer health algorithms to the app store. These analysis algorithms, packaged as services, fuse data from microdatabases and related data interfaces (e.g. weather data) to assess the health of the transformers. A Solution SME contributes the asset health app and the microdatabase for asset events and maintenance. In the regional tier, fleet health analytics are offered as app services by Technical SMEs for power transmission and distribution grids. These algorithms blend and analyze data produced in local operations tiers with data from regional microdatabases and related data interfaces, to support improved fleet planning and optimization decisions. Another Solution SME contributes the fleet health microdatabase and the fleet health app for user interactions and orchestration in regional operations. These SMEs may use a tool like SME Workbench [27] to publish their components to the app store.

Configuration activities encompass the key tasks listed in the following subsections. These tasks can be performed iteratively or in bulk, and configurations may be revisited as deployment environments evolve over time.

### A. Engineering

Configuring a composable application starts with integrating its elements into the tiers of the ecosystem instance for execution. In our example, the transformer health app services, the asset health app and microdatabase are deployed across multiple local operation center tiers. To simplify this configuration step, Solution SMEs use a tool like SME Workbench having suitable deployment extensions for the Local Operations tiers or for a calculation engine associated with the Operating Historian. Likewise, the Asset Health User Interface (UI) can be deployed or installed to the Local tiers from the app store.

Industry-specific Fleet Health algorithms for planning and optimization are obtained from the app store and deployed as app services in the regional tier. The Fleet Health UI app and microdatabase are deployed from the app store to the regional tier. Note that our IoTSP framework does not require that all tier components come from the app store; some may be installed conventionally.

Each local operations microdatabase shares its data to the regional tier by replicates its data with a corresponding microdatabase in the regional tier according to its EULA policy. Figure 10 illustrates these connections for our example.

### B. App service and microdatabase discovery

An SME determines which capabilities are available locally or via work requests with other tiers by configuring the microdatabases, app services and apps; what types of data are available and how they are obtained. In our example, the Solution SME who performs Engineering identifies the data for the Transformer Health services and microdatabases from the Operating Historian, Local Substation Computer, and weather data broker. Alternately, the Solution SME configures Fleet Planning and Optimization services to dynamically discover tiers and components via a web service, then configures the Fleet Health app to connect to those microdatabases and to its own conventional database. When apps are designed as extensible product line architectures that can automatically discover available app services or microdatabase plugins at runtime, manual configuration of these aspects can be eliminated.

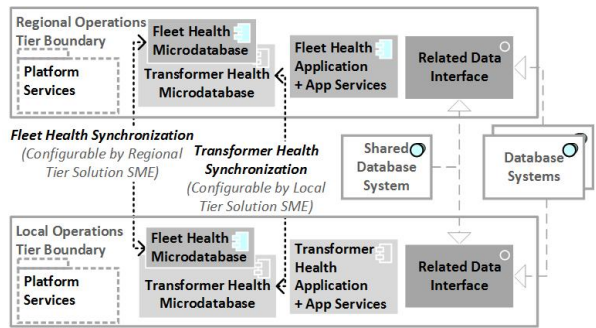

Figure 10 Applications, Services, and Microdatabases in Transformer Asset Health and Fleet Health Example Applications

*C. Asset identity discovery*

App services blend data from multiple heterogeneous data sources. In an asset-based application, different asset identities may be used in different data sources. A Solution SME configures the apps and app services with suitable data bindings within and across tiers, correctly identifying the same asset in each data source system. Tools can facilitate mapping and reconciling asset identities across the data sources.

For example, Transformer Health app service inputs may include nameplate data (e.g. vendor, voltage class, date placed into service) accessed via an owner-assigned tag number from the Asset Registry. As-installed data (e.g. location and topology), operational (e.g. actual voltage load), condition monitoring (e.g. dissolved gas analysis), and as-maintained data (e.g. last degassing service) may be identified in the Operating Historian by a grid ID. Weather data from a data broker may key on the transformer's geographic location. As-shipped data from the Asset Vendor's data store may be identified by the serial number assigned to the transformer during manufacturing.

For Fleet Health services, identity mappings may be sourced via a Related Data Interface connection to a CIR/OOR-capable Asset Registry in the Enterprise tier, or obtained in the Regional tier via an interoperability register. Note that asset identity mappings and data fusion are more complicated when assets are movable. Over long industrial lifetimes, even large power transformers might be taken out of service for refurbishment, then redeployed in another location or as a spare. The monitoring equipment associated with assets is more portable. For instance, since dissolved gas monitors can be fairly expensive, only some power transformers may not have them, while distribution transformers will not, and grid operators may move them from one troubled power transformer to another over time. An interoperability register can help to handle these historical gaps and temporal variabilities.

*D. Data bindings*

Once the data source systems and data types have been successfully discovered, and asset identities across systems have been mapped, configuration for the Transformer Health app service binds the configuration points so that relevant input data is provided for each transformer. The data it produces is stored with traceability and can be tracked and displayed by the Asset Health app with a suitable identity.

Data bindings for tags and signals for each transformer are sourced by a generic equipment model configured in the SCADA Historian and mirrored in the Operating Historian. Monitoring, failure, and maintenance inputs are bound directly to other microdatabases in the local tier. Other bindings, e.g. for obtaining relevant weather data from an external provider, are handled dynamically via a Related Data Interface, or predefined in SME Workbench.

After the Transformer Health app service has calculated updated health scores and profiles and generated recommended alternatives for actions which can be taken to improve transformer performance, this data is stored to the Transformer Health microdatabase (in the same tier as the app service) via the configured data bindings. The new asset diagnostic and reliability data are now viewable by Local Operations personnel using the Asset Health app. For our example Fleet Health app, fleet and enterprise data configuration points can be bound to microdatabases in the Regional Operations tier, schemas for conventional in-tier databases, or external data brokers.

*E. Data distribution*

A key task involves replication and synchronization of data generated by the apps and app services. In our example, the Industrial Owner wants the newly calculated local transformer diagnostic and reliability data to be available to regional app services and apps, i.e. Fleet Health, Planning, and Optimization. Accordingly, the data stored in the local Transformer Health microdatabase is replicated with a corresponding microdatabase in the regional tier, based on the EULA policy configuration determined by the owner of the Local tier. Then it is combined with other data from microdatabases within the regional tier or with externally sourced data, e.g. data from transformer vendors on fleet reliability under diverse environmental conditions.

*F. Tuning*

After configuration and testing of these applications for specific deployment instances, tuning by an Integration SME can deliver more accurate results by considering environmental conditions in which the units in the transformer fleet, sub-fleets, or families operate (e.g. heat, humidity, corrosiveness, operational demands and patterns). This tuning modifies algorithm parameters used in the services, e.g. updated fleet reliability curves or settings, which are persisted in a microdatabase within the tier.

VI. COMPLEMENTARY STANDARDS

Interoperability is a key architecture quality for the Industrial Internet, and especially for our IoTSP. Standards pave the way for successful collaboration and integration between vendors. None of the ongoing IIoT standards initiatives cover all aspects of the ecosystem. Therefore, we expect it a union of standards will enable composable Industrial Internet applications.

## A. Big Data and the Internet of Things

The IIoT consortia do not aspire to create new standards, but seek to influence the specifications. One channel for this influence is the US National Institute of Standards and Technology (NIST), with standards creation as one of their primary missions. The International Organization for Standardization (ISO) and International Electrotechnical Commission (IEC) collaborate on emerging standards through the joint Information Technology Technical Committee (JTC 1), and here we see the NIST influence. Two JTC 1 working groups in particular are considering complementary aspects of IIoT: WG9 – Big Data and WG10 – Internet of Things.

The convenor of WG9 is the digital data advisor for NIST, allowing the scope of WG9 to be revealed through his activities. The NIST work is organized as a reference architecture for Big Data [25]. The draft architecture identifies five functional components: System Orchestrator, Data Provider, Big Data Application Provider, Big Data Framework Provider, and Data Consumer, assumed to be deployed with cloud technologies. There is a significant focus on security and privacy in the ecosystem but less consideration for the sources of the data or integration with industrial processes.

Less is known about the WG10 draft reference architecture, but in general the work covers the cyber-physical systems (CPS) aspects missing from the WG9 scope. NIST provides some visibility as their CPS work is input to WG10. The architecture [26] is organized in layers, identified as Physical, Cyber and Internet tiers. These are supported by a detailed taxonomy of IoT entities and their relationships.

## B. Machine to Machine Communication

Efforts to formulate standards for device connectivity, especially in the telecom industry, pre-date ideas for IIoT. The oneM2M functional architecture [5] is organized in layers: Application, Common Services, and Network Services. The services use an adapter pattern for interoperability and highlight the need to configure, troubleshoot and upgrade the services.

## C. Open O&M

Open O&M includes MIMOSA's Software Interoperability Model [30] referring to three architectural dimensions. The Business Context describes what business the system addresses, the Information Context identifies what knowledge the system contains, and the Technology Configuration catalogs what technology the system contains. Key MIMOSA principles are the harmonization of reference data from the engineering and procurement phase with the execution environment, and consideration of the business and information contexts or dimensions as well as the technology context/dimension.

MIMOSA's most recent joint venture is the Open Industrial Interoperability Ecosystem (OIIE), which is pursuing a solutions architecture framework based on standards for cross-industry, system-of-systems interoperability in enterprise architectures [30]. MIMOSA and PCA (POSC Caesar Association) jointly released the PCA-MIMOSA Reference Architecture Framework for Integrated Engineering and Operations [28] in December 2013. Its Technology Configuration dimension describes system lay-out and structure per the Purdue Enterprise Reference Architecture, which is roughly analogous to the ISA-95 level model.

## D. Analysis

Our vision is to provide the same IoTSP service interfaces and capabilities in all tiers: cloud, enterprise, local and device. Each tier is realized with its own technology stack, but the APIs can be common to enable composable applications. Our architecture approach is to use each of the standards as the foundations and then provide an abstraction layer on top to supply the IoTSP interfaces and interactions. For example, in the cloud the Big Data Application Provider implements data collection and analytics, and the Big Data Framework Provider implements data exchange and notification. Nothing prevents the same Big Data concepts from being applied at the enterprise level, but our experience shows this is unlikely. For the enterprise the Cyber tier, including SCADA and Operating Historians, implements data collection and analytics, and factory automation systems in the same tier implement data exchange and notification. The underlying protocols are different from the cloud but with the IoTSP abstractions it is possible to design and compose applications that operate in both environments.

In the Open O&M approach, instead of a variety of point-to-point connections, each element in a system-of-systems is engineered to speak a shared O&M language over a shared information bus. The MIMOSA Open System Architecture for Enterprise Application Integration (OSA-EAI) [29] focuses in part on the prevalence of many independent, proprietary data repositories. A key MIMOSA principle is that the value of the data can be magnified by merging these repositories into an information "data network" that can be easily understood and utilized. Our perspective on tiers and microdatabases aligns with this principle and clarifies data ownership and management for the independent, proprietary repositories. However, we advocate data-oriented integration among replicated microdatabases, in lieu of integration via OSA-EAI bridges using a shared information bus.

MIMOSA's CIR and OOR concepts are both complementary to our IoTSP. CIR enables asset-based data fusion across microdatabases, whereas OOR enhances interoperability across applications and interrelationships among people, systems, and services. To maximize ecosystem interoperability and efficiency, interoperability register tools are implemented as shared platform services.

More advanced toolsets and platforms which provide machine-readable configuration information and semantic context, such as CASCoM [20], can enable self-configuration of composable applications. In our architecture, the information model clients for microdatabases and machine-readable EULA policies simplify configuration of bindings and access rights.The PCA-MIMOSA reference architecture framework provides reference models for Service Agreement, System Engineering, Software Interoperability, Semantic Ontology, and Standards Utility. Its Software Interoperability Characteristics include a chosen Technology Platform, Architectural Style, Programming Paradigm, Integration Mechanism, and Data Storage. These characteristics guide the mappings of the elements of a composable application to tiers and spaces.

## VII. Conclusions and Future Work

The future looks bright for Industrial Internet applications, taking the next step in industrial automation to connect and benefit from managed data sharing and integrated services including advanced analytics. Emerging standards in big data, cyber-physical systems, machine to machine communications, and interoperable configuration guide the abstractions essential to allow collaboration between the ecosystem vendors, third party providers and customers.

Our experience affirms the IIoT architecture is organized in tiers that interoperate, but continue to function when disconnected for technology or business reasons. We propose an IoTSP composable application development and management framework in response to these conditions. The framework can integrate legacy systems and provides platform services based on our novel data storage and management abstractions, and encourages application design based on the MVC pattern for interactions between microdatabases, app services and apps.

Deploying the applications, services and data stores is not complete without configuration. SMEs benefit from convenient selection of components and effective tools for combining and customizing the features, without need for advanced programming skills. The desired capabilities include engineering of algorithms, discovery and binding of assets and data, and deployment validation and tuning.

With these concepts and vision in place the real work begins. Our next steps are to create a reference implementation on each of the tiers and demonstrate the functionality, interactions and expected architecture qualities.